\documentclass{article}
\usepackage{spconf,amsmath,graphicx,amssymb, amsfonts}
\usepackage{xcolor}
\usepackage{bm}
\usepackage{multirow}
\usepackage[ruled,vlined]{algorithm2e}
\usepackage{url}
\usepackage{comment}
\usepackage{float}
\usepackage[shortlabels]{enumitem}
\usepackage{caption}
\usepackage{subcaption}

\DeclareMathOperator*{\argmin}{arg\,min}

\newcommand{\R}{\mathbb{R}}

\newcommand{\myx}{\bm{x}}

\newcommand{\fulldim}{p} 
\newcommand{\subdim}{d} 

\newcommand{\subspace}{\mathcal{W}} 

\newcommand{\pen}{\gamma} 

\newcommand{\noise}{\varepsilon} 
\newcommand{\noisevec}{\bm{\varepsilon}} 
\newcommand{\noisemat}{E} 
\newcommand{\unibasis}{Z} 

\newcommand{\rowind}{r}

\title{Subspace Change-point Detection via Low-Rank Matrix Factorisation}
\name{Euan T. McGonigle$^{\star}$ \qquad Hankui Peng$^{\dagger}$\thanks{Euan T. McGonigle is supported by Leverhulme Trust Research Project Grant RPG-2019-390. Hankui Peng is supported by Aviva Plc.}}
  
  \address{$^{\star}$ School of Mathematics, University of Bristol \qquad 
      $^{\dagger}$ DAMTP, University of Cambridge}

\begin{document}
\ninept
\maketitle
\begin{abstract}
Multivariate time series can often have a large number of dimensions, whether it is due to the vast amount of collected features or due to how the data sources are processed. Frequently, the main structure of the high-dimensional time series can be well represented by a lower dimensional subspace. As vast quantities of data are being collected over long periods of time, it is reasonable to assume that the underlying subspace structure would change over time. In this work, we propose a change-point detection method based on low-rank matrix factorisation that can detect multiple changes in the underlying subspace of a multivariate time series.  Experimental results on both synthetic and real data sets demonstrate the effectiveness of our approach and its advantages against various state-of-the-art methods.
\end{abstract}
\section{Introduction}\label{sec:intro}

Multiple areas of research, including signal processing and statistical learning, have demonstrated the benefit of taking advantage of low-dimensional structures (such as subspaces) in high-dimensional data. It is commonly the case that a high-dimensional data set can be well explained by a lower dimensional representation. Utilising low-dimensional subspaces can improve the efficiency of the computation and the interpretability of the analysis.

With the explosion of big data in recent years, data are routinely being collected over large timescales. In such cases, it is unlikely that the properties of the data stay the same over the entire observed period. Change-point detection (CPD) methods aim to detect locations within a time series at which the underlying properties abruptly change. It enables the segmentation of a time series into stationary segments, allowing for a parsimonious representation under less strict assumptions than that of stationarity. Applications of CPD methods can be found in a wide variety of areas, such as cyber security \cite{heard2016dynamic}, biology \cite{jewell2020fast}, and seismology \cite{li2018high}. In particular, subspace CPD methods have been applied to motion segmentation and capture data~\cite{vidal2005generalized, jiao2018subspace}, and seismic event detection~\cite{xie2020sequential}.

Due to changes in subspace manifesting as changes in the covariance of the data, there are parallels between subspace CPD and CPD methods in other areas. Firstly, subspace CPD is related to research on factor analysis of high-dimensional time series, particularly prominent in econometrics \cite{forni2000generalized}. A change in subspace can be viewed as a change in the loading matrix of a dynamic factor model. For methods to detect multiple changes in factor models, see for example \cite{barigozzi2018simultaneous,bai2020estimation} and the references therein. 
Secondly, subspace CPD can be interpreted as a special case of the change in covariance problem, as noted in \cite{xie2020sequential}. 
For CPD methods for detecting changes in the covariance matrix of high-dimensional data, see for example \cite{cho2015multiple,avanesov2018change, steland2020testing} and references therein.

There exists a vast amount of literature on the use of subspace analysis within clustering~\cite{liu2012robust, elhamifar2013sparse, you2016scalable}; however comparatively less attention has been given in the area of time series. Several authors, including \cite{doukopoulos2008fast} and \cite{narayanamurthy2018provable}, aim to track a subspace that evolves slowly over time. \cite{xie2012change} assumes that the observed time series lie close to a low-dimensional manifold, which can fluctuate slowly before a change-point, and change abruptly at a change-point. \cite{jiao2018subspace} considers an online setting where the aim is to detect a change in subspace as quickly as possible whilst additional data are being collected. \cite{xie2020sequential} also performs online subspace change-point detection, detecting an emerging subspace in the data. 
Although there are several existing methods for online detection of subspace change-points, less attention has been paid to the offline setting. The offline approach is more appropriate for applications such as motion segmentation, where we retrospectively detect all prominent changes after having observed the full time series data. 

In this paper, we propose a subspace change-point detection method, termed~\textit{SubCPD} for short, for detecting multiple changes in an offline setting. We incorporate a low-rank matrix factorisation approach within SubCPD for the estimation of low-dimensional subspaces in a given high-dimensional time series data set.    
By utilising an efficient matrix factorisation procedure, we avoid the need for computationally costly eigen-decompositions as in \cite{jiao2018subspace} and \cite{xie2020sequential}. Moreover, directly exploiting the low-dimensional structure of data enables subspace change-points to be more readily identified when compared to factor or covariance-based approaches. Through experiments on synthetic and real data, we demonstrate that our method compares favourably against state-of-the-art methods.
.

\section{Problem Formulation}
Let $\{ \myx_t \}_{t=1}^n$ denote a multivariate time series with $\myx_t \in \mathbb{R}^{\fulldim}$ for $t=1, \ldots , n$. We assume that the data lie in an unknown lower dimensional subspace~$\subspace$, i.e.
\begin{equation}
\myx_t = W_t \bm{s}_t + \noisevec_t,
\end{equation}
where $\{ \noisevec_t\in \mathbb{R}^{\fulldim} \}_{t=1}^n $ is a Gaussian random noise vector with independent entries $\noise_{\rowind t}$, $\rowind = 1, \ldots , \fulldim$. 
The whole array of variables $\{ \noise_{\rowind t} \}$ satisfies $\mathbb{E} (\noise_{\rowind t} ) =0$ and $\mathbb{E} (\noise_{\rowind t}^2 ) =\sigma^2$.
The columns of $W_t\in\R^{\fulldim \times \subdim}$ form an orthonormal basis of the underlying subspace with $\text{dim} (W_t)=\subdim$, and $\bm{s}_t\in\R^{\subdim}$ is the low-dimensional representation of the signal~$\myx_{t}$. 

If a set of $K$ changes have occurred, $\mathcal{T}=\left\{\tau_{i} \right\}_{i=1}^{K}$, the underlying subspace basis $W_t$ undergoes a change at each of the $K$ \emph{change-points} $1 \leq \tau_1 < \tau_2 < \ldots < \tau_K \leq (n-1)$. The data segment between any pair of adjacent change-points $\tau_{i}$ and $\tau_{(i+1)}$ can be summarised well with a common subspace basis~$Z_{i}\in\mathbb{R}^{p\times\subdim}$,~i.e.\ 
\begin{equation*}
W_{\tau_{i}+1} = \ldots = W_{\tau_{(i+1)}} := Z_i, \  \forall \ 0 \leq i \leq K,
\end{equation*}
with the convention that $\tau_0:=0$ and $\tau_{(K+1)}:= n$. Our aim is to estimate the locations of the $K$ changes $\tau_1, \ldots, \tau_K$, and the $(K+1)$ subspace bases $Z_0, \ldots, Z_K$. 

If no change has occurred, i.e.\ $K=0$, then the data can be well represented with a common subspace basis~$Z_0\in\mathbb{R}^{p\times\subdim}$.
Let us denote by $X \in \mathbb{R}^{p \times n}$ the entire data matrix; $S \in \mathbb{R}^{d \times n}$ the concatenated columns of the low-dimensional representations $\bm{s}_{1}, \ldots, \bm{s}_{n}$; and $\noisemat\in\mathbb{R}^{p\times n}$ the concatenated columns of the noise vectors $\noisevec_{1}, \ldots, \noisevec_{n}$. Then the change-free time series data set can be represented in the following matrix form 
\begin{equation}
X = \unibasis S + \noisemat,    
\end{equation}
where the subscript in $Z_{0}$ is dropped here and in the rest of the paper when there is no ambiguity.

\section{Proposed Method}

In this section, we propose a subspace change-point detection method that utilises a matrix factorisation technique for effectively and efficiently estimating the subspace structure given a data set. For the ease of discussion, let us first consider the case where we have no change in subspace and the data set $X$ contains only one underlying subspace structure $\unibasis$.
We wish to find $\widehat{\unibasis}$ and $\widehat{S}$ that approximate the data matrix $X$ well, which leads us to the following matrix factorisation problem:
\begin{equation}\label{mat-fac}
(\widehat{\unibasis},\widehat{S}) = \min_{\substack{\unibasis \in \mathbb{R}^{p \times d}, \\S \in \mathbb{R}^{d \times n}}} \{  L (X, \unibasis S ) + \lambda R(\unibasis,S)  \},
\end{equation}
where $L(X, \unibasis S )$ is a loss function that measures how well certain $\unibasis$ and $S$ approximate the data matrix $X$, $R(\unibasis,S)$ is a regularisation term for enforcing desired properties on $\unibasis$ and $S$, and $\lambda$ is a regularisation parameter. 

\subsection{Solving the Matrix Factorisation Problem}

Our choice of the loss function and regularisation term in~\eqref{mat-fac} are driven by two main properties that are desirable for the subspace CPD problem at hand. 
Firstly, we wish to find $\unibasis$ and $S$ that approximate data well. This can be achieved by measuring the sum of squares between the points (${\myx_{t}}$)s and each of their approximations (${Z\bm{s}_{t}}$)s. Concretely, the sum of squares error translates into the following loss function
\begin{equation}
L(X, \unibasis S ) = \|  X- \unibasis S  \|_{F}^2,
\end{equation}
where $||  \cdot  ||_{F}$ is the Frobenius norm.
Secondly, since $X$ is assumed to have a low-dimensional subspace structure, we wish to enforce the low-rankness in the approximation $\unibasis S$. This is because the nuclear norm is often used to approximate matrix rank, and it can optimally recover low-rank matrices under certain conditions~\cite{recht2010guaranteed}. Concretely, we wish to solve the following matrix factorisation problem
\begin{equation}\label{eq:main_prob}
(\widehat{\unibasis}, \widehat{S})=\min_{\substack{\unibasis \in \mathbb{R}^{p \times d}, \\S \in \mathbb{R}^{d \times n}}} \{  \|  X- \unibasis S  \|_{F}^2 + \lambda \| \unibasis S \|_{*}  \}.
\end{equation}

It is worth noting that each of the sub-problems in~\eqref{eq:main_prob} of optimising over $\unibasis$ or $S$ is a convex problem, whilst the other variable is kept fixed. This allows for the use of an efficient alternating minimisation strategy called block coordinate descent~\cite{xu2013block}. It optimises the main objective function by solving each of the convex sub-problems iteratively. Although the alternating minimisation strategy does not guarantee convergence to a local minimum, good results can be obtained empirically with trivial initialisations of $\unibasis$ and~$S$~\cite{haeffele2014structured}. 

\subsection{Detecting a Single Change}\label{single-change}
We first consider the setting where there is a single subspace change at time $\tau$, before extending to multiple changes in Section \ref{multiple-changes}. Denote by $X_{s:e}$ the data matrix comprised of the data columns of $\myx_{s}$ up to $\myx_{e}$. 
At each candidate change location $k$, we evaluate the fit of the matrix factorisation to the left and right of change location $k$. Using the criterion in~\eqref{eq:main_prob}, we obtain the overall loss from the two matrix factorisations as 
\begin{align}\label{test-stat}
 \nonumber T(k) =& || X_{1:k}- \widehat{Z}_l \widehat{S}_l ||^2_{F}+ || X_{(k+1):n}- \widehat{Z}_r \widehat{S}_r ||^2_{F} +\\
& \lambda || \widehat{Z}_l \widehat{S}_l  ||_* + \lambda || \widehat{Z}_r \widehat{S}_r ||_*, 
\end{align}
where $\widehat{Z}_l$ and $\widehat{S}_l$ are defined as the solutions to~\eqref{eq:main_prob} on the data segment $X_{1:k}$ to the left of the location $k$. Similarly, $\widehat{Z}_r$ and $\widehat{S}_r$ are the solutions on the data segment $X_{(k+1):n}$ to the right of $k$. 
Intuitively, if a change is present at the true change-point location~$\tau$, then the best low-rank approximation to the data involves two matrix factorisations; to the left and right of~$\tau$. Therefore, lower values of the test statistic~\eqref{test-stat} correspond to greater evidence of a change in subspace. The estimator of the change location is given by 
\begin{equation}
 \hat{\tau} = \argmin_{1 \leq k \leq n } T(k),   
\end{equation}
which denotes the point in time that minimises the overall loss in~\eqref{test-stat}. 

We conclude that a change-point is detected at time $\hat{\tau}$ if the following criterion
\begin{equation}\label{model-selection}
L (X_{1:\hat{\tau}}, \widehat{Z}_l \widehat{S}_l) + L(X_{(\hat{\tau}+1):n}, \widehat{Z}_r \widehat{S}_r)  +\pen <  L(X, \widehat{\unibasis} \widehat{S}) 
\end{equation}
is satisfied, where $\widehat{\unibasis}$ and $\widehat{S}$ consist the solution to~\eqref{eq:main_prob} on the entire data set $X$. Here, $\pen$ is an additional penalisation term to protect against overfitting the number of change-points. We specify the penalisation term $\pen$ to be $\pen=\mu \log n$, among other potential choices of the penalty~\cite{arlot2019kernel}.

It can be viewed as a model selection step analogous to the Bayesian Information Criterion (BIC); see for example \cite{kaul2019efficient}. In summary, we detect a change if the penalised loss at~$\hat{\tau}$ is less than the loss assuming no change.

\subsection{Detecting Multiple Changes}\label{multiple-changes}
To detect multiple changes in subspace, we can use the Binary Segmentation (BS) algorithm~\cite{scott1974cluster}. The BS algorithm initially searches the entire data set for a single change-point, by solving the optimisation problem in~\eqref{eq:main_prob} to calculate the test statistic \eqref{test-stat}. If a change-point is detected using criterion \eqref{model-selection}, the data are split into two sub-segments defined by the detected change-point. Then, the procedure is recursively repeated on subsequent sub-segments until no further changes are detected. The algorithmic form of our proposed multiple subspace change-point detection method utilising binary segmentation, referred to henceforth as SubCPD, is summarised in Algorithm~\ref{algo-1}. 

\begin{table*}[t]
\centering
\resizebox{2\columnwidth}{!}{
\begin{tabular}{cc|cc|cc|cc|cc|cc|cc|cc|cc|cc}
  &  &  \multicolumn{6}{c|}{A} &  \multicolumn{6}{c|}{B} &  \multicolumn{6}{c}{C}   \\ \cline{3-20}
      &        &  \multicolumn{2}{c|}{SubCPD}     & \multicolumn{2}{c|}{FAC}    & \multicolumn{2}{c|}{SBS} &  \multicolumn{2}{c|}{SubCPD}     & \multicolumn{2}{c|}{FAC}    & \multicolumn{2}{c|}{SBS} &  \multicolumn{2}{c|}{SubCPD}     & \multicolumn{2}{c|}{FAC}    & \multicolumn{2}{c}{SBS}   \\
   $p$ & $d$ & TNC & VM & TNC & VM & TNC & VM & TNC & VM & TNC & VM & TNC & VM & TNC & VM & TNC & VM & TNC & VM \\ \hline
   20 & 2 & \bf{999} & \bf{0.998} & 521 & 0.880 & 884 & 0.932 & \bf{999} & \bf{0.998} & 588 & 0.889 & 940 & 0.943 & \bf{998} & \bf{0.985} & 350 & 0.838 & 222 & 0.819 \\
       & 4 & \bf{1000}  & \bf{1.000} & 910 & 0.921 &  895 & 0.937 &  \bf{969} & \bf{0.998} & 926 & 0.925 &  933 & 0.943 &  \bf{996} & \bf{0.994} & 759 & 0.888 & 321 &0.844  \\
    & 6 & \bf{951} & \bf{0.998} & 909 & 0.915 & 708 & 0.912 & \bf{994} & \bf{1.000} & 935 & 0.916 & 756 & 0.916 & \bf{828} & \bf{0.988} & 803 & 0.887 & 252 & 0.833 \\ \hline
      50 & 4 & \bf{1000} & \bf{1.000} & 917 & 0.922 & 893 & 0.939 & \bf{1000} & \bf{1.000} & 904 & 0.923   &  924 & 0.944 & \bf{999} & \bf{0.994} & 653 & 0.875 & 323 & 0.840 \\
    & 7 & \bf{1000} & \bf{1.000} & 988 & 0.935 & 878 & 0.939  & \bf{1000} & \bf{1.000} & 990 & 0.939 & 895 & 0.943 & \bf{1000} & \bf{0.998} & 898 & 0.900 & 410 & 0.861 \\
    & 10 & \bf{1000} & \bf{1.000} & 989 & 0.934 & 806 & 0.925 & 846 & \bf{0.990} & \bf{981} & 0.934 & 840 & 0.932 & \bf{974} & \bf{0.998} & 830 & 0.889 & 460 & 0.864  \\ \hline
      100 & 5 & \bf{1000} & \bf{1.000} & 920 & 0.922 & 813 & 0.933 & \bf{1000} & \bf{1.000} & 929 & 0.924 & 832 & 0.940 & \bf{997} & \bf{0.998} & 575 & 0.865 & 467 & 0.843  \\
    & 10 & \textbf{1000}  & \bf{1.000} & 989 & 0.940 & 779 & 0.932 & \bf{1000} &\bf{1.000} & 984 & 0.940 & 795 & 0.937 & \bf{996} & \bf{0.998}  & 895 & 0.899 & 576 & 0.867 \\
    & 15 & \bf{987} & \bf{0.999} & 966 & 0.934 & 790 & 0.929 & \bf{994} & \bf{1.000} & 979 & 0.935 & 768 & 0.932 & 853 & \bf{0.989} & \bf{922} & 0.904 & 612 & 0.874  \\ 

  \end{tabular}}
\caption{Detection comparisons for varying $p$ and $d$ in noise scenarios from A to C. We report the number of replications that return the true number of change-points (TNC), and the average V-measure over all realisations. The best performing method in each scenario is shown in bold.}
\label{alt-table}
\end{table*}

\begin{algorithm}[htbp!]
 \KwIn{Data matrix $X_{s:e}$, regularisation parameter $\lambda$, change penalty $\mu$}
\textbf{Initialisation:} $s=1, e = n, i=1, \mathcal{T}=\emptyset$\\ 
    \For{$k \in \{s,\ldots , e\}$}{
       Calculate $T(k) \leftarrow L(X_{s:k}) + L(X_{(k+1):e})$ according to~\eqref{test-stat}\\
    }
    $\hat{\tau}_{i} \leftarrow \argmin_{s \leq k \leq e} T(k) $ \\
     \If{~\eqref{model-selection} is satisfied}{
     Add $\hat{\tau}_{i}$ to the set of estimated change-points $\mathcal{T}$\\
      SubCPD$(s,\hat{\tau},  \lambda, \mu)$ \\
      SubCPD$(\hat{\tau}+1,e,  \lambda, \mu)$ \\
     $i \leftarrow i+1$
    }
 \KwOut{The set of estimated change-points $\mathcal{T}$.}
 \caption{Subspace Change-Point Detection (SubCPD)}
 \label{algo-1}
\end{algorithm}

\section{Numerical Results}
In this section, we provide practical guidelines for the implementation of SubCPD. We demonstrate its strong performance on synthetic data sets by comparing against state-of-the-art methods. Lastly, we illustrate the value of SubCPD on motion capture data segmentation.

\subsection{Experimental Set-up \& Practical Considerations}

We first detail the parameter settings we use for obtaining the experimental results on synthetic and real data sets. Then we discuss some practical considerations around subspace CPD in general and computational speed.  

\textbf{Choice of regularisation parameter $\lambda$ in~\eqref{test-stat}.} In line with \cite{haeffele2014structured}, we set $\lambda = \hat{\sigma}/2$, where $\hat{\sigma}$ is a robust estimate of the noise $\sigma$ calculated using the median absolute deviation. If the data are serially correlated, then this could be replaced with an estimate of the long-run variance.

\textbf{Choice of $\mu$ for protection against overfitting.} When the true number of change-points $K$ is known in advance, we can run SubCPD up to $K$ changes with no penalty term. The next change is chosen as the location that maximally reduces the loss function. 

To set $\mu$ in a fully automatic manner, we can use the ``slope heuristic" \cite{baudry2012slope} \cite{arlot2019kernel}. The idea is that the penalty term can be estimated by evaluating the loss function for $\tau$ number of changes for values of $\tau$ much larger than $K$. We first run the method with no penalty up to a prescribed number of changes $\tau_{\text{max}}$, similar to the known $K$ case. Then, we perform a linear regression of the loss assuming $\tau$ change-points against $\log (n/\tau)$ for $\tau \in [0.6 \tau_{\text{max}}, \tau_{\text{max}}]$. We set $\mu = -2\hat{s}$, where $\hat{s}$ is the estimated regression coefficient as in \cite{arlot2019kernel}.

\textbf{Practical considerations.} In CPD methods it is common practice to set a Minimum segment length (MSL), i.e. the minimum distance between two consecutive change-points. This helps avoid detecting spurious change-points near the boundaries and improve computation time. In line with \cite{matteson2014nonparametric}, we use $\text{MSL}=30$. 

Another general consideration for subspace CPD methods is the dimension $d$ of the subspace structure. If $d$ is known a priori this can be used in the procedure. Otherwise, we calculate the ratio of consecutive ordered eigenvalues of the sample covariance matrix on an initial portion of the data, and choose $d$ to be the value that minimises this quantity, as described in \cite{lam2012factor}. 

Finally, we discuss the computational capability of our proposed SubCPD method. 
By design, our proposed SubCPD method scales linearly with the length of the time series data.  This is due to the need to perform the minimisation step in~\eqref{eq:main_prob} at each time point to calculate the overall loss~\eqref{test-stat}. To improve upon this, we can instead evaluate \eqref{test-stat} over an equally spaced grid of $\log n$ locations, rather than every time point, as in \cite{kaul2021inference}. The location of the change can be refined by evaluating \eqref{test-stat} in a neighbourhood of the estimated~$\hat{\tau}$. This reduces the number of optimisation steps from $\mathcal{O} (n)$ to $\mathcal{O} (\log n)$, at a small expense of the detection capability.

\begin{figure*}[t!]
\centering
\begin{subfigure}[b]{0.33\textwidth}
 \centering
 \includegraphics[width=\textwidth, height=.5\textwidth, trim={5mm 80mm 5mm 80mm}]{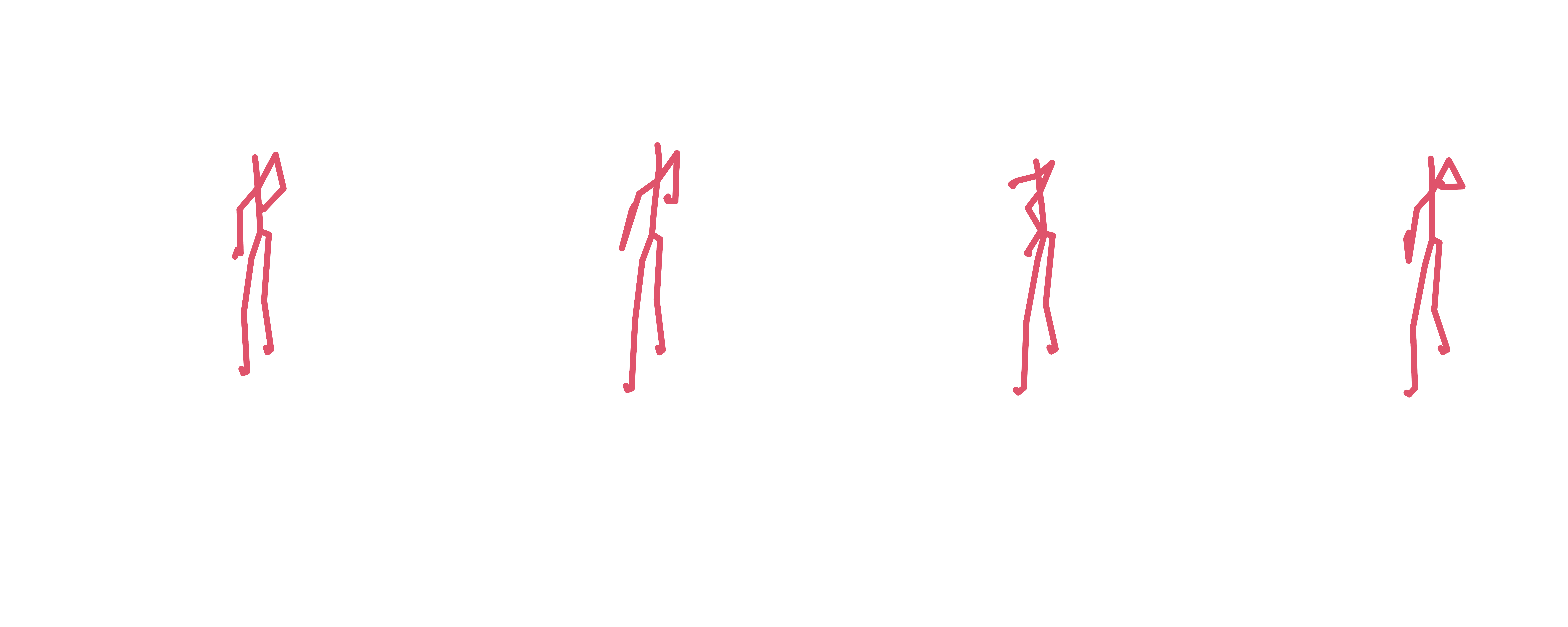}
 \vspace{-1cm}
 \caption{Punching (with right arm).}
 \label{fig:motion_a}
\end{subfigure}
\hfill
\begin{subfigure}[b]{0.33\textwidth}
 \centering
 \includegraphics[width=\textwidth, height=.5\textwidth, trim={5mm 80mm 100mm 100mm}]{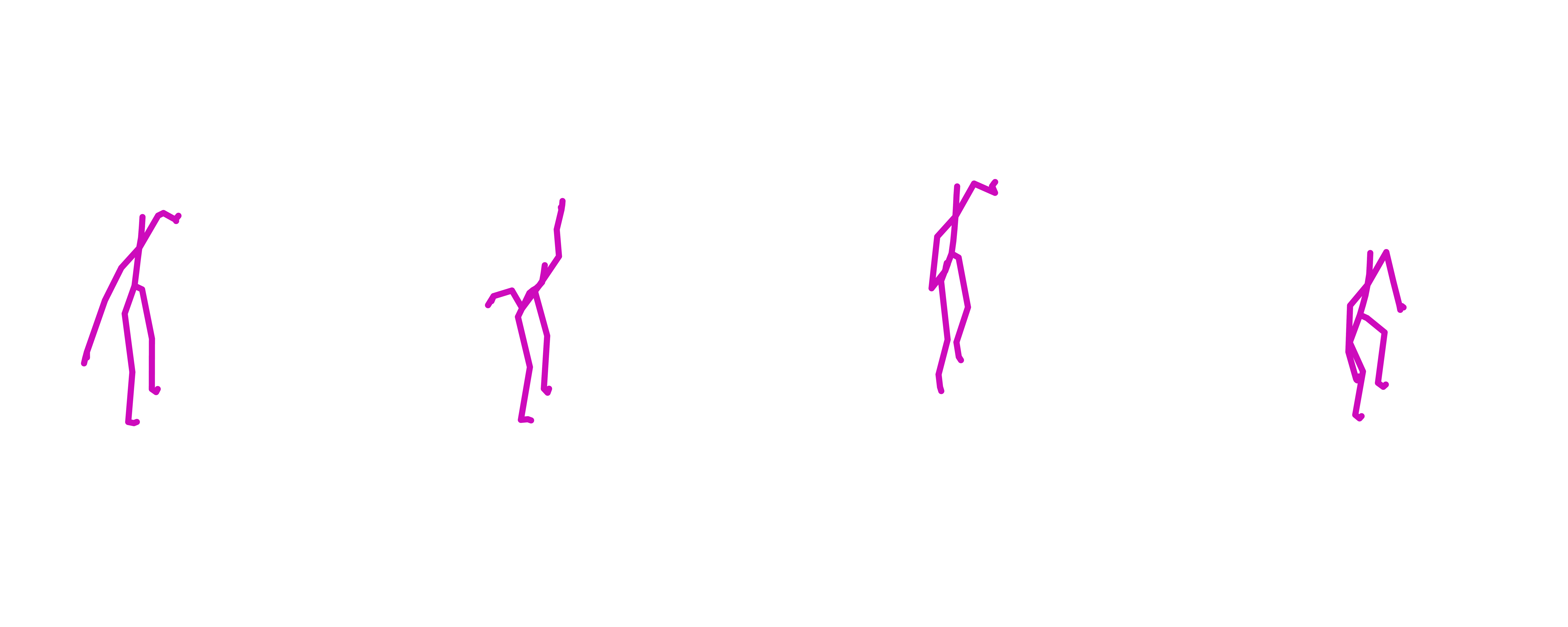}
 \vspace{-1cm}
 \caption{Jumping (with two legs).}
 \label{fig:motion_b}
\end{subfigure}
\hfill 
\begin{subfigure}[b]{0.33\textwidth}
 \centering
 \includegraphics[width=\textwidth, height=.5\textwidth, trim={5mm 80mm 5mm 80mm}]{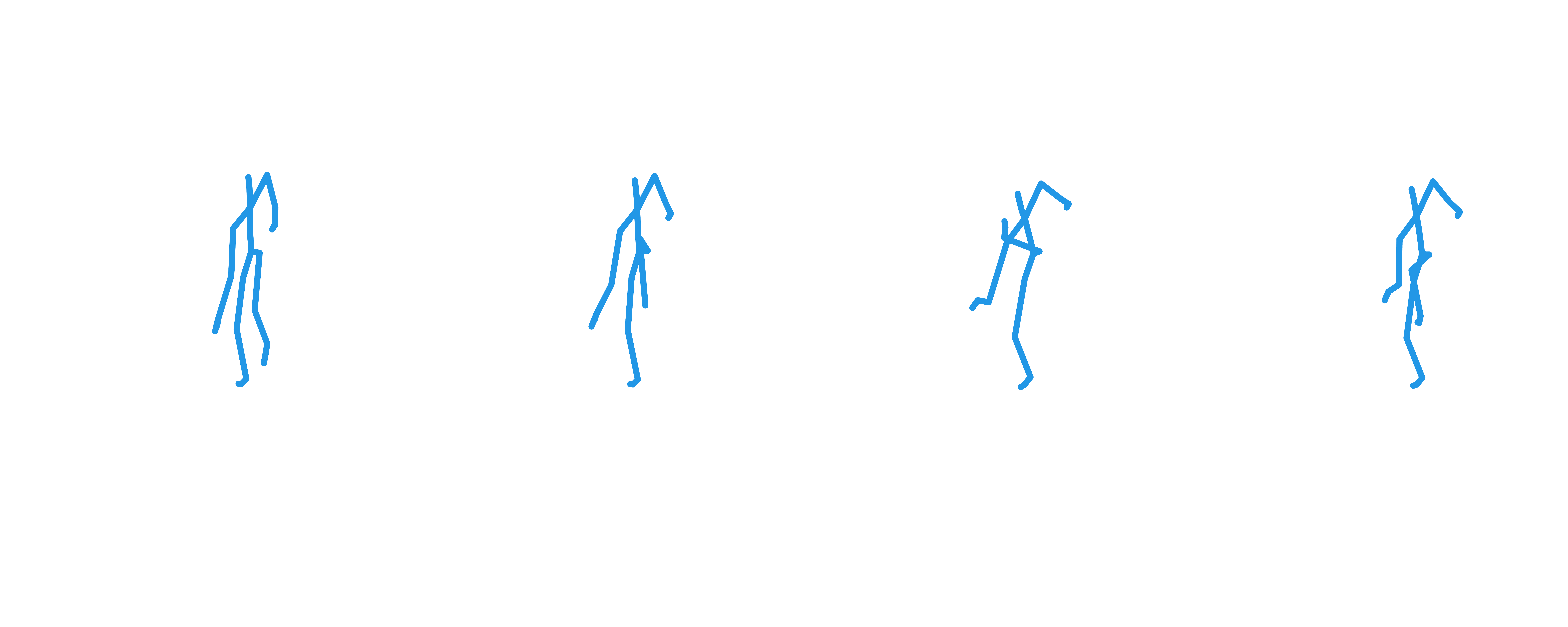}
 \vspace{-1cm}
 \caption{Kicking (with left leg).}
 \label{fig:motion_c}
\end{subfigure}
\caption{Three series of still frames taken over the course of trial No.\ 1 of subject No.\ 86. Each of the three series consists of four frames that show a sequence of body movements from one physical activity (e.g.\ punching, jumping, kicking).}
\label{motion-plot}
\end{figure*}

\begin{figure}[htbp!]
\centering
\includegraphics[width =\columnwidth]{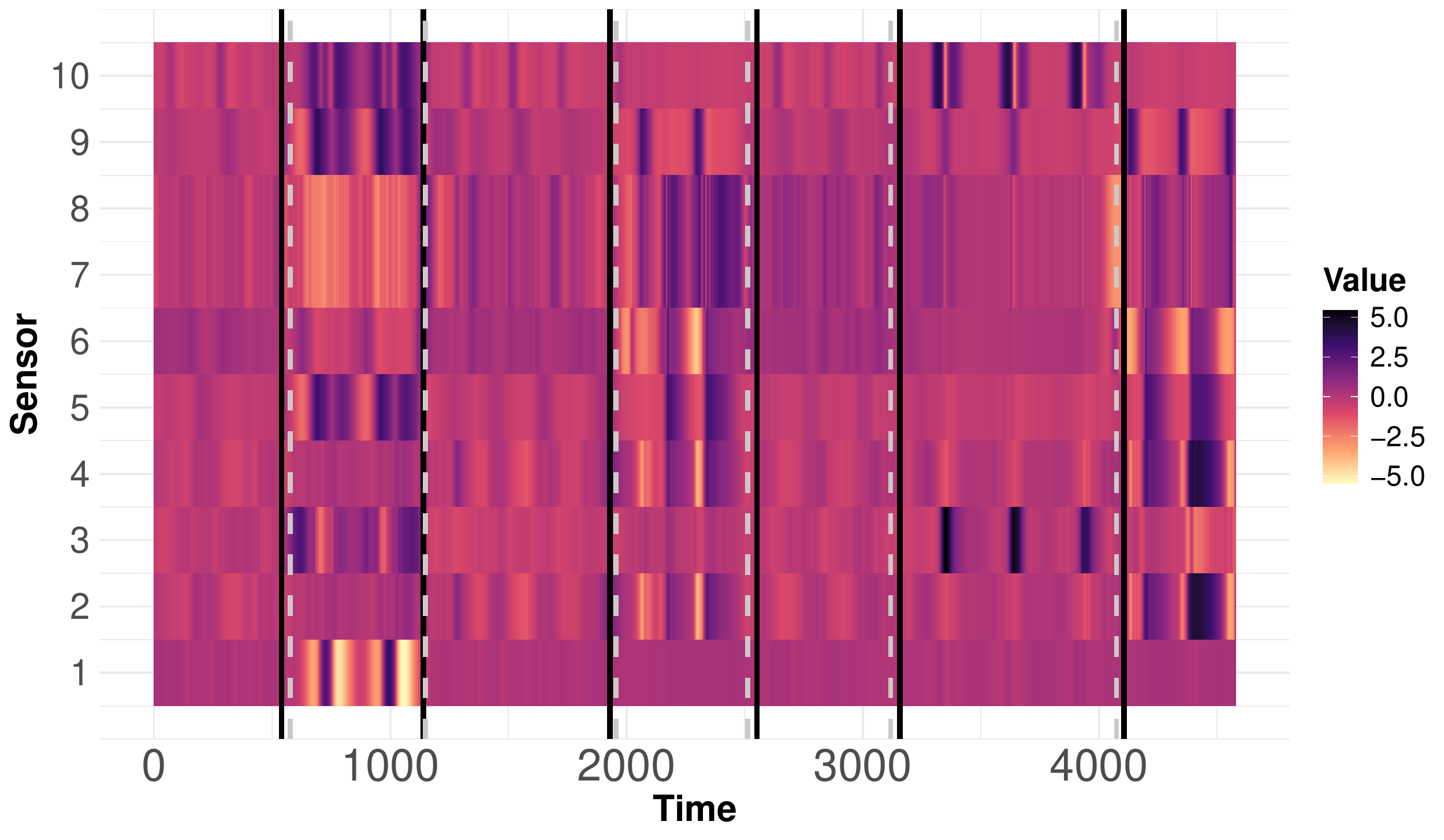}
\caption{Heatmap of a 10 variable subset of the motion capture data set. True change-points shown in solid black lines, detected change-points by SubCPD in dashed grey.}
\label{mocap-plot-1}
\end{figure}

\subsection{Synthetic Data}

We simulate 1000 replications of time series of length $n=500$  while varying the values of $p$ and $d$. The initial subspace basis $Z_0$ is generated by orthonormalising a matrix of independent and identically distributed (IID) standard normal variables. Data are simulated following~\cite{jiao2018subspace}, where the subspace basis $Z_{(i+1)}$ is generated such that the distance between subspaces $D (Z_i, Z_{(i+1)})^2 = d - || Z_i^{\mathsf{T}} Z_{(i+1)} ||_F  = \Delta^2$ for some $\Delta>0$. We set $\Delta = \sqrt{d}/2$, with the maximum distance being $\sqrt{d}$. Each replication has 4 change-points, given by $\tau = (100,200,300,400)$. To assess the robustness of the method to noise, we examine three scenarios of $\noise_t$:
\begin{enumerate}[(A)] 
\item  IID $\noise_t \sim N(0,0.005)$ as in \cite{jiao2018subspace};
\item the $\noise_t$ follow the AR(1) process $\noise_t = 0.7 \noise_{t-1} + Y_t$ where $Y_t$ is zero-mean normal such that $\text{Var}(\noise_t)=0.005$, i.e. serially correlated noise; and 
\item IID $\noise_t \sim N(0,0.05)$ i.e. scenario with higher level of noise.
\end{enumerate}  

To the best of our knowledge, we are the only CPD method that exploits subspace structure in an offline setting.  
As such, we compare SubCPD to the factor-based approach of \cite{barigozzi2018simultaneous} (FAC) and covariance-based sparsified binary segmentation (SBS)~\cite{cho2015multiple}. We report the total number of the 1000 simulations that return the true number of changes (TNC) and the V-measure (VM) of the resulting segmentation obtained from the detected change-points. The VM quantifies the similarity between two different segmentations of a data set and takes values on $[0,1]$~\cite{rosenberg2007v}. A larger V-measure corresponds to a more accurate segmentation of the data, with a V-measure of 1 indicating perfect segmentation.

The results are reported in Table \ref{alt-table}, with the best performing method in each setting given in bold. We see that SubCPD offers the strongest performance across almost all scenarios, both in terms of TNC and VM. This demonstrates the benefits of exploiting low-dimensional structure in a higher dimensional data set for change-point detection.

\subsection{Motion Capture Data Segmentation}
In this section, we evaluate the performance of our proposed SubCPD method on the Carnegie Mellon Motion Capture data set (MoCap) \footnote{The data set can be downloaded from~\url{http://mocap.cs.cmu.edu/}.}. It contains data recordings from 62 sensors placed at various joints on a human test subject. There are 144 human test subjects in total, and each human test subject conducts a number of trials. Within each trial, the subject goes through 6 to 12 different activities, such as walking, jumping, kicking, etc. Given the 62-dimensional time series collected from the sensor data of a certain trial, our aim is to detect the change-points of when the subject changes from one activity to another. Previous works have found that motion sensor data for a single activity can be summarised well with a 5-dimensional subspace~\cite{jiao2018subspace}. As such, it can be suitably cast as a subspace change-point detection problem.

\begin{figure}[t!]
\centering
\includegraphics[width=0.85\columnwidth]{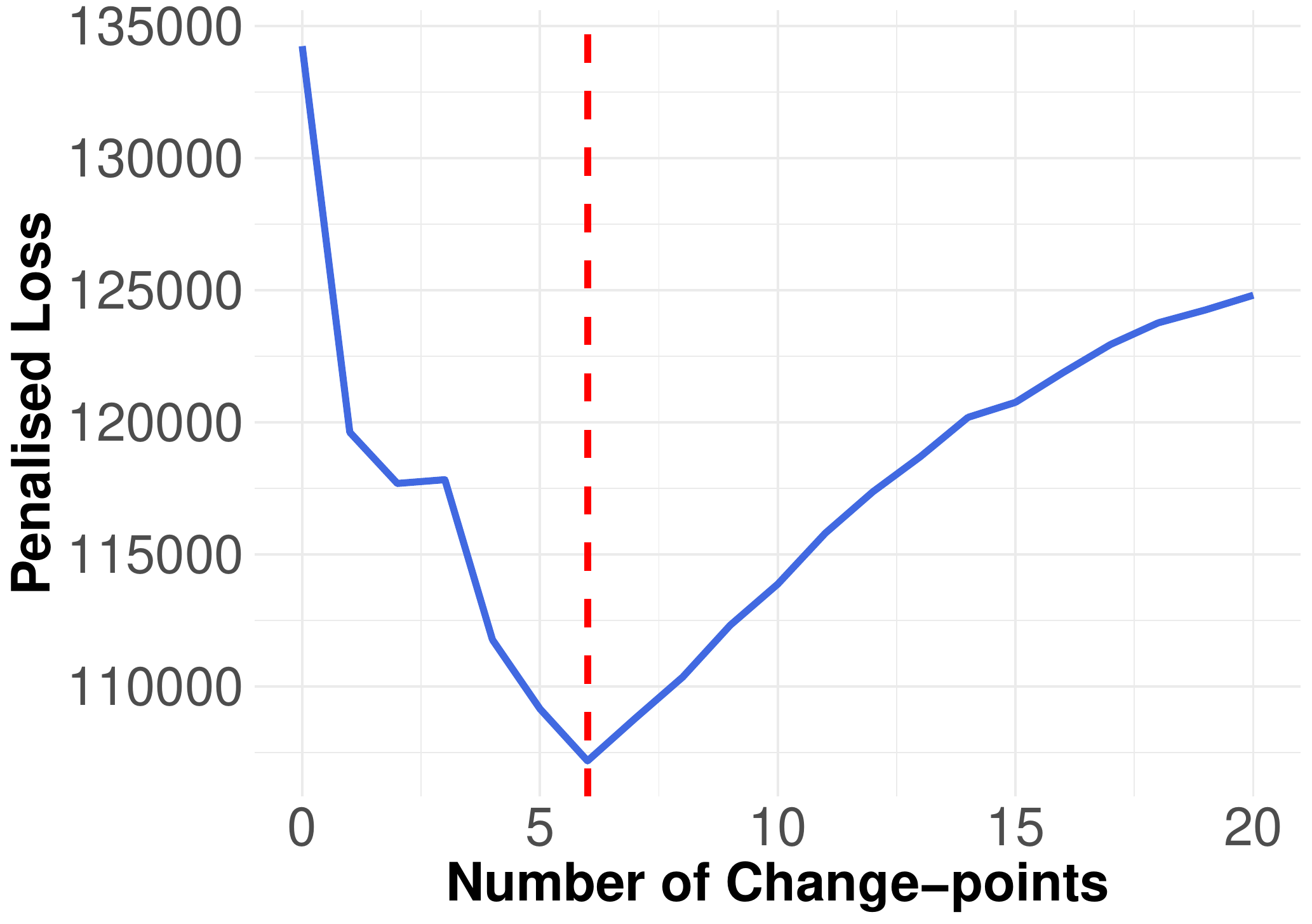}
\caption{Plot showing the penalised loss for a specified number of change-points, minimised at 6 changes.}
\label{scree-plot}
\end{figure}

We focus on trial 1 of subject 86 as an illustrative example. Fig. \ref{motion-plot} shows a series of still frames, tracking the subject's movements over the course of the trial. There are 6 changes in subspace, caused when the subject transitions between activities. Before applying SubCPD, we standardise each series by subtracting the mean and dividing by the standard deviation. We run SubCPD using the slope heuristic to estimate the penalty term $\mu$. The resulting plot of the penalised loss function over a range of change-point numbers is shown in Fig. \ref{scree-plot}, which suggests 6 changes. The results of the segmentation are shown in Fig. \ref{mocap-plot-1}, where for ease of visualisation, we plot a heatmap of a 10 variable subset of the data. The detected changes are shown in dashed line, while locations of the true changes are given in solid line. As well as detecting the correct number of changes, SubCPD also locates them accurately. 

\section{Conclusions \& Future Work}
We have proposed a subspace change-point detection method (SubCPD) that exploits the underlying low-dimensional subspace structure in a higher dimensional time series. It can efficiently and effectively detect the locations of multiple change-points under a variety of scenarios, including high-dimensional and serially correlated data. 
Through extensive synthetic experiments, we demonstrate the competitive performance of SubCPD against various state-of-the-art methods. Furthermore, we showcase the capability of SubCPD through the frequently studied motion capture data set. 

For future work, we would like to consider other matrix factorisation methods, such as \cite{jaggi2010simple}. We could also apply the method with more sophisticated algorithms for detecting multiple change-points, such as wild binary segmentation \cite{fryzlewicz2014wild}. 

\bibliographystyle{IEEEbib}
\bibliography{refs}

\end{document}